\documentclass[a4paper,12pt]{article}
\usepackage{amsmath,amssymb,epsfig}
\usepackage{setspace}
\usepackage{graphicx}
\usepackage{caption}
\usepackage{subcaption}

\newcommand{\bra}[1]{\langle #1|}
\newcommand{\ket}[1]{|#1\rangle}
\newcommand{\braket}[2]{\langle #1|#2\rangle}

\title{Action with Acceleration I: Euclidean Hamiltonian and Path Integral}
\author{\normalsize
\begin{tabular}[t]{c@{\extracolsep{3em}}c@{\extracolsep{3em}}c@{\extracolsep{3em}}c}
\large Belal E. Baaquie \\
Department of Physics, National University of Singapore \\
2 Science Drive 3, Singapore 117542, Singapore\\
phybeb@nus.edu.sg
\end{tabular}}
\date{\today}
\begin{document}
\maketitle

\begin{abstract}
An action having an \textit{acceleration} term in addition to the usual velocity term is analyzed. The quantum mechanical system is directly defined for Euclidean time using the path integral. The Euclidean Hamiltonian is shown to  yield the acceleration Lagrangian and the  path integral with the correct boundary conditions. Due to the acceleration term, the state space depends on \textit{both} position and velocity -- and hence the Euclidean Hamiltonian depends on two degrees of freedom. The Hamiltonian for the acceleration system is non-Hermitian and can be mapped to a Hermitian Hamiltonian using a  similarity transformation; the matrix elements of this unbounded transformation is explicitly evaluated. The mapping fails for a critical value of the coupling constants. 
\end{abstract}

\section{Introduction}
The action with acceleration arises in many diverse fields; the Euclidean action and path integral, studied  by Hawking and Hertog  \cite{hawking}, arises in quantum gravity and D-brane dynamics;  Bender and Mannheim have extensively studied the problem of ghost states in quantum mechanics using the Minkowski action \cite{mannheim2}; the action appears is the study of mathematical finance, describing the time dependence of financial instruments \cite{bebcup2}, and the Euclidean path integral has been studied extensively in \cite{bebcyeqt}, \cite{cythesis}; the action is as an example of  higher derivative Lagrangians  \cite{PhysRevD.41.3720}; the path integral for pseudo-Hermitian Hamiltonians has been studied starting from the propagator for the theory \cite{jones} and \cite{rjrivers}.

Euclidean time is defined by the analytic continuation of Minkowski time $t_M=-i\tau$, where $\tau$ is Euclidean time. The degree of freedom $x$ does not change in going from Minkowski to Euclidean time but Minkowski velocity $v_M$ picks up an extra factor of $i$ since it is related to Euclidean velocity $v$ by the following\footnote{The minus in the definition of Euclidean velocity follows from $v=iv_M$, which is consistent with $\tau=it_M$.}
\begin{equation}
\label{cnstrxv}
v=-\frac{dx}{d\tau}=i\frac{dx}{dt_M}=iv_M  
\end{equation}
The non-Hermitian Euclidean Hamiltonian and path integral are well behaved with a single analytic continuation of time, and no further complexification of the degrees of freedom is required. In contrast, an analytic continuation of the degrees of freedom is required for the Minkowski case, as discussed in \cite{mannheimmain}.

\section{The Feynman path integral}
Consider the Euclidean time Lagrangian with  acceleration given by
\begin{equation}
\mathcal{L}=-\frac{\gamma}{2}(\frac{d^2x}{dt^2})^2-\frac{\alpha}{2}(\frac{dx}{dt})^2-\Phi(x)
\end{equation}
with the acceleration action for finite Euclidean time $\tau$ given by
\begin{equation}
\mathcal{S}[x]=\int_0^\tau dt \mathcal{L}
\end{equation}

The Feynman path integral for finite Euclidean time is given by
\begin{eqnarray}
\label{eq:partition}
&&\mathcal{K}(x_f,\dot{x}_f;x_i,\dot{x}_i)=\int DX~e^{\mathcal{S}[x]}\Big{|}_{(x_i,\dot{x}_i;x_f,\dot{x}_f)}\\
&&\int DX=\tilde{\mathcal{N}}\prod_{t=0}^\tau \int_{-\infty}^{+\infty}dx(t)
\end{eqnarray}
where $\mathcal{\tilde{N}}$ is a normalization constant. The paths have the following four boundary conditions
\begin{eqnarray}
\label{bchiger}
&&x(0)=x_i;~~\frac{dx(0)}{dt}=\dot{x}_i~~ \text{initial position and velocity}\\
&&x(\tau)=x_f;~~\frac{dx(\tau)}{dt}=\dot{x}_f~~ \text{final position and  velocity}
\end{eqnarray}

Since the path integral given Eq. \ref{eq:partition} is quadratic, it can be evaluated exactly using the classical solution. Let $x_c(t)$ be the classical solution given by
\begin{equation}
\label{cleqnmotion}
\frac{\delta \mathcal{S}[x_c]}{\delta x(t)}=0
\end{equation}
that satisfies the following boundary conditions
\begin{eqnarray}
\label{bcclass}
&&x_c(0)=x_i;~~\frac{dx_c(0)}{dt}=\dot{x}_i~~ \text{initial position and velocity}\\
&&x_c(\tau)=x_f;~~\frac{dx_c(\tau)}{dt}=\dot{x}_f~~ \text{final position and  velocity} \nonumber
\end{eqnarray}

Consider the following change of integration variables, from $x(t)$ to $\xi(t)$
\begin{equation}
x(t)=x_c(t)+ \xi(t)
\end{equation}
with boundary conditions for given by
\begin{eqnarray}
&&\xi(0)=0;~~\frac{d\xi(0)}{dt}=0~~ \text{initial position and velocity}\\
&&\xi(\tau)=x_f;~~\frac{d\xi(\tau)}{dt}=0~~ \text{final position and  velocity} \nonumber
\end{eqnarray}
The change of variables yields
\begin{eqnarray}
&&\mathcal{S}[x]=\mathcal{S}[x_c+\xi]=\mathcal{S}[x_c]+\mathcal{S}[\xi] \nonumber\\
\label{bcclasskerslon}
&&\int DX~e^{\mathcal{S}}=\mathcal{N}(\tau)e^{\mathcal{S}[x_c]}\\
&&\text{where}~~\mathcal{N}(\tau)=\int D\xi~e^{\mathcal{S}[\xi]}\nonumber
\end{eqnarray}
Hence, From Eqs. \ref{eq:partition} and \ref{bcclasskerslon}
\begin{eqnarray}
\label{bcclassker}
&&\mathcal{K}(x_f,\dot{x}_f;x_i,\dot{x}_i)=\mathcal{N}(\tau)e^{\mathcal{S}[x_c]}
\end{eqnarray}
The evolution kernel given in Eq. \ref{bcclassker} has been evaluated explicitly in \cite{hawking} and \cite{cythesis} by solving for the classical solution $x_c(t)$ and then obtaining $\mathcal{S}[x_c]$ and $\mathcal{N}(\tau)$.

As can be directly verified from the classical action  $\mathcal{S}[x_c]$, the classical solution $x_c(t)$ given by Eq. \ref{cleqnmotion} yields \textit{another} equally valid classical solution $\tilde{x}_c(t)$ given by  the following 
\begin{eqnarray}
\label{eq:symmpartitionadd}
&&\tilde{x}_c(t)=x_c(\tau-t)
\end{eqnarray}
with boundary conditions, from Eq. \ref{bcclass}, given by
\begin{eqnarray}
&&\tilde{x}_c(0)=x_c(\tau)=x_f;~~\frac{d\tilde{x}_c(0)}{dt}=-\frac{dx_c(\tau)}{dt}=-\dot{x}_f\\
&&\tilde{x}_c(\tau)=x_c(0)=x_i;~~~\frac{d\tilde{x}_c(\tau)}{dt}=-\frac{dx_c(0)}{dt}=-\dot{x}_i\\
 \nonumber
 &&~~~~~~~\Rightarrow~~~~~\mathcal{S}[x_c]=\mathcal{S}[\tilde{x}_c]
\end{eqnarray}
Hence
\begin{eqnarray}
\label{bcclassadd}
\mathcal{S}_c[x_f,\dot{x}_f;x_i,\dot{x}_i]=\mathcal{S}[x_i,-\dot{x}_i;x_f,-\dot{x}_f]
\end{eqnarray}

The classical action $S_c$ is given in Eq. \ref{eq:action}; the solution is seen to have the symmetry given in Eq. \ref{bcclassadd}.

The evolution kernel, from Eqs. \ref{bcclassker} and \ref{bcclassadd},  has the following symmetry
\begin{eqnarray}
\label{eq:symmevolker}
&&\mathcal{K}(x_f,\dot{x}_f;x_i,\dot{x}_i)=\mathcal{K}(x_i,-\dot{x}_i;x_f,-\dot{x}_f)
\end{eqnarray}

\section{Euclidean Hamiltonian and Path Integral}
The derivation of $\mathcal{K}(x_f,\dot{x}_f;x_i,\dot{x}_i)$ was done entirely in terms of the coordinate degree of freedom $x(t)$ and made no reference to any other degrees of freedom; in particular, the velocity degree of freedom $v(t)$ did not appear in the path integral derivation.

One would like to interpret the evolution kernel $\mathcal{K}(x_f,\dot{x}_f;x_i,\dot{x}_i)$ as the \textit{probability amplitude} for a transition from an initial to a final state vector. Such an interpretation of course needs both a state space and a Hamiltonian.

Based on the boundary conditions given in Eq. \ref{bchiger}, it can be seen that the state space has to have \textit{two independent degrees of freedom}, corresponding to the two initial conditions given by the initial position $x$ and velocity $\dot x$.  Hence, the  state space $\mathcal{V}$ of non-Hermitian Hamiltonian is taken to have \textit{two} degrees of freedom, namely a position $x$ and  a velocity $v$ degree of freedom . The Hamiltonian has to be chosen in such a manner that the velocity degree of freedom $v$ is \textit{constrained} to be equal to the velocity $\dot x$ of the coordinate degree of freedom $x$.

Consider two independent degrees of freedom $x$ and $v$. The completeness equation for the basis states are given by
\begin{eqnarray}
\label{eq:complete}
\mathcal{I}=\int_{-\infty}^{+\infty} dx dv \ket{x,v}\bra{v,x}\\
\langle x,v|x',v'\rangle=\delta(x-x')\delta(v-v') \nonumber
\end{eqnarray}

A state space representation of the evolution kernel $\mathcal{K}(x_f,\dot{x}_f;x_i,\dot{x}_i)$ is derived. It will be shown that  the evolution kernel is closely related to the \textit{probability amplitude} of going,  in time $\tau$, from the initial state $\ket{x_i,v_i}$ to the final state $\bra{x_f,v_f}$  and is given by
\begin{equation}
\label{kerstasp}
\mathcal{K}_S(x_f,v_f;x_i,v_i)=\bra{x_f,v_f}e^{-\tau H}\ket{x_i,v_i}
\end{equation}
A similar definition is adopted for the path integral of a non-Hermitian Hamiltonian in  \cite{Kandirmaz}.
 
It remains to be seen as to what is the precise relation of the probability amplitude $\mathcal{K}_S(x_f,v_f;x_i,v_i)$ defined using the state space and Hamiltonian to the probability amplitude $\mathcal{K}(x_f,\dot{x}_f;x_i,\dot{x}_i)$ defined using the path integral. In particular, as it stands in Eq. \ref{kerstasp}, the initial $v_i$ and final velocity $v_f$ have no relation with the coordinate degree of freedom $x$. The Hamiltonian has to implement a constraint to set the initial and final state in Eq. \ref{kerstasp} to have the same as the boundary conditions given in Eq. \ref{bchiger} for $\mathcal{K}(x_f,\dot{x}_f;x_i,\dot{x}_i)$.

The Hamiltonian, for infinitesimal time $\tau=\epsilon$, is given by the Dirac-Feynman formula as follows
\begin{equation}
\bra{x,v}e^{-\epsilon H}\ket{x',v'}=\mathcal{C}(\epsilon)e^{\epsilon\mathcal{L}(x,x';v,v')}
\end{equation}
where $\mathcal{C}(\epsilon)$ is a normalization constant that depends only on $\epsilon$. The discrete time Lagrangian is given by
\begin{equation}
\label{dislagrang}
\mathcal{L}(x,x';v,v')=-\frac{\gamma}{2}\left(\frac{\dot x -\dot x'}{\epsilon} \right)^2-\frac{\alpha}{2}\left(\frac{x-x'}{\epsilon} \right)^2 -\frac{1}{2}[\Phi(x)+\Phi(x')]
\end{equation}

The Minkowski Hamiltonian for the action acceleration has been obtained by Bender and Mannheim \cite{mannheimmain}, \cite{mannheim2}; they have shown that Hamiltonian and state space for Minkowski time is well behaved, but requires an analytic continuation of the degree of freedom. 

The analytic continuation to Euclidean time of the Minkowski Hamiltonian yields a Euclidean Hamiltonian given by 
\begin{equation}
H=-\frac{1}{2\gamma}\frac{\partial^2}{\partial v^2}-v\frac{\partial}{\partial x}+\frac{1}{2}\alpha v^2+\Phi(x)
\label{eq:Hamiltonian1}
\end{equation}
The term $-v\partial/\partial x$ is noteworthy since it does not depend on any coupling constant; this term  \textit{constrains} the degrees of freedom and finally leads to the constraint that $v=-dx/dt$, as required by Eq. \ref{cnstrxv}; the constraint due to the term $-v\partial/\partial x$  in the Hamiltonian has a transparent realization in the path integral formulation and discussed in Section \ref{sec:changebc}.

Note that $H$ is not Hermitian since
\begin{equation}
H^\dagger=-\frac{1}{2\gamma}\frac{\partial^2}{\partial v^2}+v\frac{\partial}{\partial x}-\frac{\alpha}{2}v^2+\Phi(x) \ne H
\end{equation}

The transition probability amplitude is given by defining $\epsilon=\tau/N$ and inserting $N-1$ complete set of states given in Eq.~\ref{eq:complete}. Hence, for boundary conditions given by $x_0=x_i, v_0=v_i; x_N=x_f, v_N=v_f$, the transition amplitude is given by
\begin{align}
&\mathcal{K}(x_f,v_f;x_i,v_i)=\bra{x_f,v_f}e^{-\tau H}\ket{x_i,v_i}\nonumber\\
\label{eq:delta1}
&~~~~=\displaystyle\prod_{n=1}^{N-1}\int dx_n dv_n \displaystyle\prod_{n=1}^{N}\bra{x_n,v_n}e^{-\epsilon H}\ket{x_{n-1},v_{n-1}}\nonumber\\
&~~~~=\displaystyle\prod_{n=1}^{N-1}\int dx_n \bra{x_N,v_N}e^{-\epsilon H}\ket{x_{N-1},v_{N-1}}\nonumber\\
&~~~~~~~~~~~~\times\left[ \displaystyle\prod_{n=1}^{N-1} \int dv_n \bra{x_n,v_n}e^{-\epsilon H}\ket{x_{n-1},v_{n-1}} \right]
\end{align}
The reason for putting the term $\bra{x_N,v_N}e^{-\epsilon H}\ket{x_{N-1},v_{N-1}}$ outside the $\int dv_n$ integrations is because, as will be seen in Eq. \ref{keerepsilon}, this term does not depend on $v_{N-1}$.

The differential operator $H$ given in Eq. \ref{eq:Hamiltonian1}, for $x_n=x,v=v_n;~x'=x_{n-1},v'=v_{n-1}$,  yields the following\footnote{Note that the Hamiltonian acts on the dual basis states $\bra{x,v}$.}
\begin{align}
&\bra{x,v}e^{-\epsilon H}\ket{x',v'}=e^{-\epsilon H (x,\frac{\partial}{\partial x};v,\frac{\partial}{\partial v})}\braket{x}{x'}\braket{v}{v'}\nonumber\\
&=e^{-\epsilon H}\int \frac{dp}{2\pi}\frac{dq}{2 \pi} e^{ip(x-x')+iq(v-v')}\nonumber\\
&=\displaystyle \int\frac{dp}{2\pi}\frac{dq}{2 \pi} e^{-\frac{\epsilon}{2\gamma}q^2+iq(v-v')-\frac{\epsilon\alpha}{2}v^2}
e^{ip(x-x'+\epsilon v)-\epsilon\Phi(x)}\nonumber\\
\label{eq:propagator}
&=\mathcal{C}\delta(x-x'+\epsilon v)\exp\{-\frac{\gamma}{2\epsilon}(v-v')^2-\frac{\epsilon\alpha}{2}v^2-\epsilon\Phi(x)\}
\end{align}
Hence Eq. \ref{eq:propagator} yields the following
\begin{align}
\label{keerepsilon}
&\bra{x_n,v_n}e^{-\epsilon H}\ket{x_{n-1},v_{n-1}}=\mathcal{C}\delta(x_n-x_{n-1}+\epsilon v_n)\exp\{\epsilon\mathcal{L}_n\}
\end{align}
The $\delta$-function depends only on $v_n$ and results in constraining $v_n$ on reaching a boundary of the path integral.

Eq. \ref{eq:propagator} yields the remarkable result that, as mentioned earlier, the term $v\partial/\partial x$ in the Hamiltonian yields a \textit{constraint} $\delta(x-x'+\epsilon v)$ on the degree of freedom $v$ so that it is constrained to be the velocity, namely $v=-dx/d\tau$.

The appearance of the $\delta-$function in Eq. \ref{keerepsilon} yields, in the integrand of the path integral, the following constraint 
\begin{align}
\label{eq:delta21}
&\delta(x_n-x_{n-1}+\epsilon v_n)~~\Rightarrow v_n=-\frac{x_n-x_{n-1}}{\epsilon}\\
\label{eq:delta2}
&\Rightarrow ~~
\lim_{\epsilon\to 0} ~~v_n=-\frac{dx_n}{d\tau}
\end{align}

The Lagrangian $\mathcal{L}_n$, from Eqs. \ref{eq:propagator} and \ref{keerepsilon}, is given by
\begin{align}
&\mathcal{L}_n=-\frac{\gamma}{2\epsilon^2}(v-v')^2-\frac{\alpha}{2}v^2-\frac{1}{2}[\Phi(x)+\Phi(x')]\nonumber\\
\label{epslag}
&=-\frac{\gamma}{2}\left(\frac{v_n -v_{n-1}}{\epsilon} \right)^2-\frac{\alpha}{2}\left(\frac{x_n -x_{n-1}}{\epsilon} \right)^2-\frac{1}{2}[\Phi(x_n)+\Phi(x_{n-1})]
\end{align}

The path integral and Lagrangian that appears in Eq.~\ref{eq:partition} makes no reference to the integration over the velocity variables. Hence, all the velocity integrations $\int dv_n$ need to be carried out in order that one obtains the  expression in Eq.~\ref{eq:partition}.   Remarkably enough, all the $\int dv_n$ integrations can be done \textit{exactly} due to the $\delta-$ function that appears in Eq.~\ref{keerepsilon}.

Eqs. \ref{eq:delta1} and \ref{keerepsilon}  yield the following velocity path integral
\begin{align}
&\displaystyle\prod_{n=1}^{N-1}\int dv_n \bra{x_n,v_n}e^{-\epsilon H}\ket{x_{n-1},v_{n-1}}\nonumber\\
\label{eq:HL}
&~~~~~~~~=\mathcal{C}^{N-1}\displaystyle\prod_{n=1}^{N-1}\int dv_n \delta(x_n-x_{n-1}+\epsilon v_n)\exp\{\epsilon\mathcal{L}_n\}
\end{align}
\section{Change of boundary conditions} \label{sec:changebc}
The four boundary conditions given in Eq. \ref{bcclass} are solely in terms of the position degree of freedom $x(t)$ whereas the boundary conditions given in Eq. \ref{kerstasp}, the defining equation for $\mathcal{K}_S(x_f,v_f;x_i,v_i)$ is given in terms of final and initial positions and velocities $x_f,v_f;x_i,v_i$ respectively.

The initial and final time steps in the path integral need to be examined carefully to see how the initial and final velocity, $v_0$ and $v_N$ respectively, can be expressed solely in terms of the position degree of freedom. 

Note the integrand of Eq. \ref{eq:HL}, for $n=1$, yields, from Eq. \ref{epslag}, the following
\begin{align}
&\int dv_1\delta(x_1-x_{0}+\epsilon v_1)\exp\{\epsilon\mathcal{L}_1\} \nonumber\\
&\mathcal{L}_1=-\frac{\gamma}{2}\left(\frac{v_1 -v_{0}}{\epsilon} \right)^2-\frac{\alpha}{2}\left(\frac{x_1 -x_{0}}{\epsilon} \right)^2-\frac{1}{2}[\Phi(x_1)+\Phi(x_{0})]
\end{align}

On performing the $\int dv_1$ integration, the delta function constrains $v_1=-(x_1 -x_{0})/\epsilon$; hence, $\mathcal{L}_1$ has the following value
\begin{align}
\label{vivolag}
&\exp\{\epsilon\mathcal{L}_1\}=\exp\Big\{-\frac{\gamma}{2\epsilon}\left(\frac{x_1 -x_{0}}{\epsilon}+v_{0} \right)^2-\frac{\alpha}{2\epsilon}\left(x_1 -x_{0} \right)^2+O(\epsilon)\Big\}\nonumber\\
&~~~~~~~=\exp\Big\{-\frac{\gamma}{2\epsilon}\left(\frac{x_1 -x_{0}}{\epsilon}+v_{0} \right)^2-\frac{\alpha\epsilon}{2}v_0^2+O(\epsilon)\Big\}\nonumber\\
& ~~~~~~~=\mathcal{C}(\epsilon) \delta(x_1-x_{0}+\epsilon v_0) +O(\epsilon)=\mathcal{C}(\epsilon)\delta(x_1-x_{i}+\epsilon v_i)
\end{align}
where  $x_0=x_i$ and $v_0=v_i$ are the initial position and velocity.

The final time boundary term for the action yields, using the final velocity $v_N=v_f$, the following
\begin{align*}
&\bra{x_N,v_N}e^{-\epsilon H}\ket{x_{N-1},v_{N-1}}=\mathcal{C}(\epsilon)\delta(x_N-x_{N-1}+\epsilon v_N)\exp\{\epsilon \mathcal{L}_N\}\\
&~~~~~~~~~~~~~~~~~~~~~~~~~~~~~~~~~~=\mathcal{C}(\epsilon)\delta(x_f-x_{N-1}+\epsilon v_f)\exp\{\epsilon \mathcal{L}_N\}
\end{align*}
since  $x_N=x_f$ and $v_N=v_f$.

Collecting all the results yields the discrete time path integral for the transition probability expressed solely in terms of the co-ordinated degrees of freedom, namely
\begin{eqnarray}
&&\mathcal{K}_S(x_f,v_f;x_i,v_i)=\bra{x_f,v_f}e^{-\tau H}\ket{x_i,v_i}\nonumber\\
\label{discretepieuclid}
          &&=\mathcal{\tilde{C}}\displaystyle\prod_{n=1}^{N-1}\int dx_n \delta(x_f-x_{N-1}+\epsilon v_f)\delta(x_1-x_{i}+\epsilon v_i) \exp\{\epsilon \sum_{n=1}^N\mathcal{L}_n\}
\end{eqnarray}
where $\mathcal{\tilde{C}}$ is a normalization.

The path integral over the velocity degrees of freedom yields, in addition to the expected acceleration action, two delta functions. These delta functions are crucial in changing the boundary conditions for the path integral over the position degrees of freedom. 

The position path integral given in Eq. \ref{discretepieuclid}, due to the two delta-functions in the integrand, has \textit{four} boundary conditions for the position degree of freedom, namely $x_i,x_f$ due to boundary conditions from the initial and final state vectors and  two more boundary conditions imposed on $x_1,x_{N-1}$ due to the two delta-functions resulting from the velocity path integral;  in effect, these two delta-functions remove two integrations, namely $\int dx_1dx_{N-1}$ in the path integral given in Eq. \ref{discretepieuclid} by fixing the value of $x_1,~x_{N-1}$.

To take the continuum limit define
\begin{equation}
\dot x(t)=\frac{dx(t)}{dt}=\frac{x_n-x_{n-1}}{\epsilon}~~;~~t=n\epsilon
\end{equation}
Hence, from Eqs. \ref{vivolag} and \ref{discretepieuclid}, respectively
\begin{align}
&v_i=\frac{x_0-x_{1}}{\epsilon}\to -\frac{dx(0)}{dt}=-\dot{x}_i\\
\label{vneqxn}
&v_f=\frac{x_{N-1}-x_{N}}{\epsilon}\to -\frac{dx(\tau)}{dt}=-\dot{x}_f
\end{align}
From Eq.~\ref{eq:HL},
\begin{equation}
\epsilon \mathcal{L}_n=-\frac{\gamma}{2\epsilon}(\dot x_n-\dot x_{n-1})^2-\frac{\epsilon \alpha}{2}\dot x_n^2-\epsilon \Phi(x_n)
\end{equation}
Taking the limit of $\epsilon\rightarrow 0$ using $\ddot x=[\dot x_{n}-\dot x_{n-1}]/\epsilon$ yields
\begin{equation}
\label{lagpot2}
\mathcal{L}=-\frac{\gamma}{2}\ddot x^2-\frac{\alpha}{2}\dot x^2-\Phi(x)
\end{equation}
The delta functions for the boundary values of the functional integral $\int DX$ are constraints that \textit{change the boundary conditions} on the path integral, converting the two position boundary conditions in the path integral $\int DXDV$  to four boundary conditions for the path integral $\int DX$. Hence, one obtains the following continuum result
\begin{eqnarray*}
&&\mathcal{K}_S(x_f,v_f;x_i,v_i)=\bra{x_f,v_f}e^{-\tau H}\ket{x_i,v_i}\\
          &&~~=\int DX\delta(v_i+\dot{x}_i)\delta(v_f+\dot{x}_f)e^S\Big{|}_{(x_i,x_f)}\\
          &&~~=\int DXe^S\Big{|}_{(x_i,\dot{x}_i=-v_i;x_f,\dot{x}_f=-v_f)}\\
          &&~~~=\mathcal{K}(x_i,\dot{x}_i=-v_i;x_f,\dot{x}_f=-v_f) \\
          &&\Rightarrow~~\mathcal{K}_S(x_f,v_f;x_i,v_i)=\mathcal{K}(x_f,-v_f;x_i,-v_i)
\end{eqnarray*}

Recall from Eq.~\ref{eq:Hamiltonian}, the Hamiltonian for Lagrangian in Eq. \ref{lagpot2} is given by
\begin{equation}
H=-\frac{1}{2\gamma}\frac{\partial^2}{\partial v^2}-v\frac{\partial}{\partial x}-\frac{\alpha}{2}v^2+\Phi(x)
\end{equation}

\section{Equivalent Hermitian Hamiltonian $H_0$}
To analyze $H$ in greater detail, choose the Gaussian potential
\begin{equation}
\Phi(x)= \frac{\beta}{2}x^2
\end{equation}
that yields the Hamiltonian and Lagrangian given by
\begin{eqnarray}
\label{eq:Hamiltonian}
H=-\frac{1}{2\gamma}\frac{\partial^2}{\partial v^2}-v\frac{\partial}{\partial x}+\frac{1}{2}\alpha v^2+\frac{\beta}{2}x^2~~;~~ \mathcal{L}=-\frac{1}{2}\left[\gamma\ddot x^2+\alpha \dot x^2+\beta x^2 \right]~~
\end{eqnarray}
A parametrization that is more suitable for studying the Hamiltonian and state space is given by  \cite{mannheimmain} 
\begin{equation}
\label{lagpot}
\mathcal{L}=-\frac{\gamma}{2}\left[\ddot x^2+(\omega_1^2+\omega_2^2)\dot x^2+\omega_1^2\omega_2^2x^2 \right]
\end{equation}
Note the Lagrangian is completely symmetric in parameters $\omega_1$ and $\omega_2$. 

The Hamiltonian given by
\begin{equation}
\label{hamw1w2}
H=-\frac{1}{2\gamma}\frac{\partial^2}{\partial v^2}-v\frac{\partial}{\partial x}+\frac{\gamma}{2}(\omega_1^2+\omega_2^2)v^2+\frac{\gamma}{2}\omega_1^2\omega_2^2x^2
\end{equation}

The Hamiltonian acts on a state space $\mathcal{V}$ with \textit{two} degrees of freedom, namely position position  $x$ and velocity degree of function $v$. The state vecor $\ket{\Psi}$ is given by
\begin{equation}
\ket{\Psi} \in\mathcal{V}~~;~~ \braket{x,v}{\Psi}=\Psi(x,v)
\end{equation}

\begin{figure}[h]
  \centering
  \epsfig{file=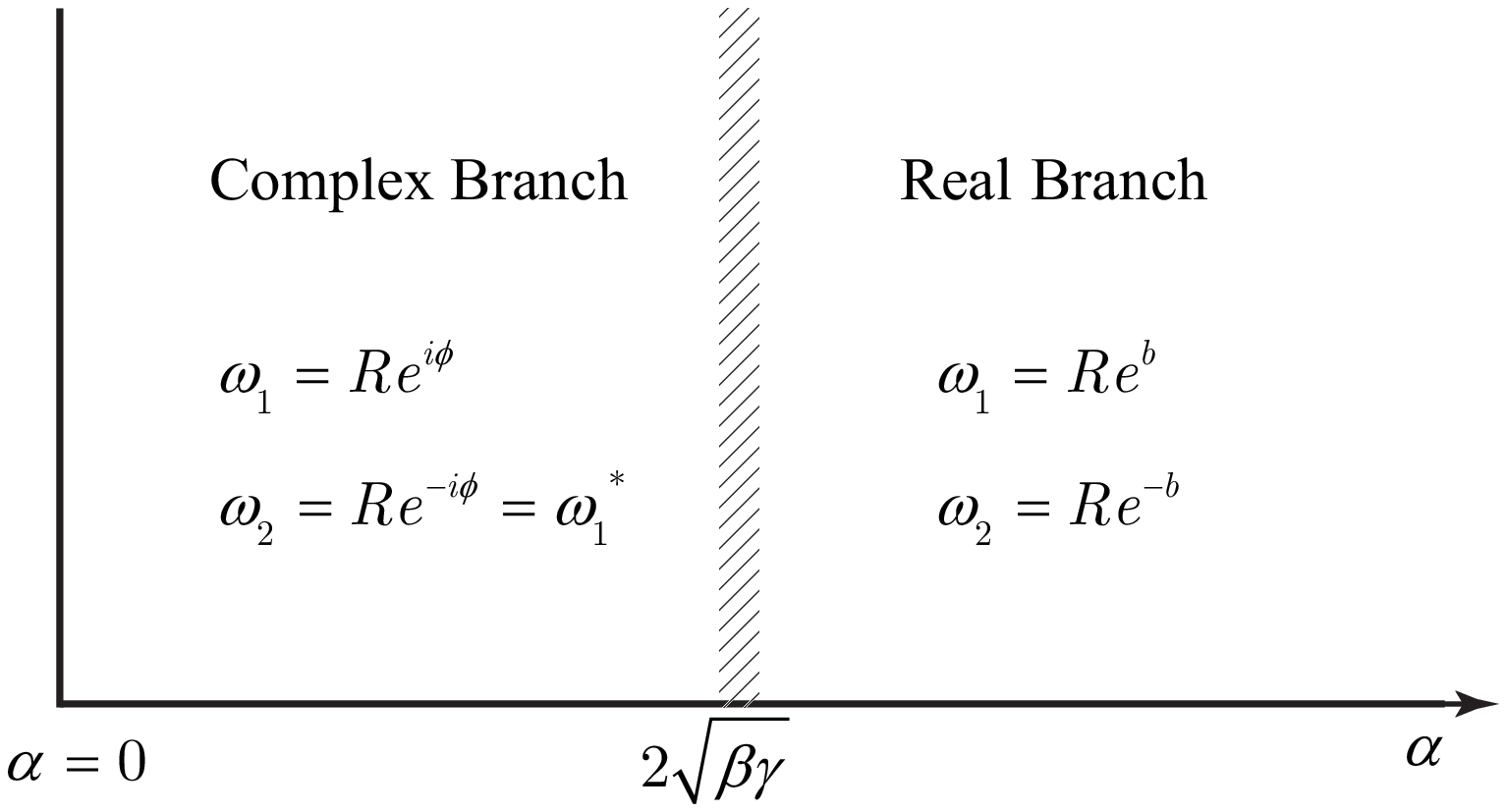, height=6cm}
  \caption{Parameter branches.}
  \label{brances}
\end{figure}

The parameters have three branches and are shown in Figure \ref{brances}.
\begin{itemize}
\item Complex branch $\alpha<2\sqrt{\beta\gamma}$.

Frequencies $\omega_1, \omega_2$ are complex.
\begin{align}
\omega_1= \omega_2^*=Re^{i\phi}~~:~~R>0,~~\phi\in[-\pi/2,\pi/2]\nonumber
\end{align}

Note $\phi\in[-\pi/2,\pi/2]$ for all $\alpha,\beta>0$ -- and this is also the range for which the path integral is convergent.

\item Real branch $\alpha>2\sqrt{\beta\gamma}$.

Frequencies $\omega_1, \omega_2$ are real and $\omega_1> \omega_2$ is chosen without any loss of generality.
\begin{align}
\omega_1= Re^{b}~~;~~\omega_2=Re^{-b}~~:~~R>0,~~b\in[0,+\infty] \nonumber
\end{align}

For the case of real $\omega_1$ and $\omega_2$, the entire parameter space is covered by choosing say $\omega_1>\omega_2$ and the roots are chosen accordingly as given below
\begin{eqnarray}
\label{omega1omega2}
&&\omega_1=\frac{1}{2\sqrt{\gamma}}\left(\sqrt{\alpha+2\sqrt{\gamma\beta}}+\sqrt{\alpha-2\sqrt{\gamma\beta}}\right)\nonumber\\
&&\omega_2=\frac{1}{2\sqrt{\gamma}}\left(\sqrt{\alpha+2\sqrt{\gamma\beta}}-\sqrt{\alpha-2\sqrt{\gamma\beta}}\right)~\\
&&\omega_1>\omega_2~~\text{for $\omega_1,~\omega_2$ real} \nonumber
\end{eqnarray}

\item Equal frequency $\alpha=2\sqrt{\beta\gamma}$.
\begin{align}
\omega_1=\omega_2~~;~~ b=0=\phi \nonumber
\end{align}
The special case of equal frequency $\omega_1=\omega_2$ is treated in detail in \cite{bebeujordan}.
\end{itemize}

A similarity transformation is obtained such that
\begin{equation}
e^{-Q/2}He^{Q/2}=H_{O}
\label{eq:replace}
\end{equation}
where the Hamiltonian $H_{O}$ is a system of two decoupled harmonic oscillators, one each for degree of freedom $x$ and $v$.

In a pioneering paper,  Bender and Mannheim \cite{mannheimmain} found the operator $Q$ for the Minkowski case; the Euclidean version of their result is given by the following  
\begin{align}
\label{Qoperdef}
Q&=axv-b\frac{\partial^2}{\partial x \partial v}
\end{align}
For the real domain where $\omega_1,~\omega_2$ are real, both coefficients $a,b$ are real and hence
\begin{align}
\label{Qherm}
Q&=Q^\dagger:~~\text{Hermitian for $\omega_1,~\omega_2$ real}
\end{align}
The definition of $Q$ continues to hold for the complex domain but  $Q$ is no longer Hermitian.
 
To obtain $H_0$,  the equation for the commutator
\begin{equation}
e^{-Q}\mathcal{O}e^{Q}=\displaystyle\sum_{n=0}^{\infty}\frac{1}{n!}\underbrace{\left[ \left[ [\mathcal{O},Q],Q\right]\ldots Q\right]}_{n-\text{fold commutator}}
\end{equation}
needs to be applied to $\mathcal{O}=x,v,\partial/\partial x, \partial/\partial v$.

To obtain the commutator, note that the $n$-fold commutator of $Q$ with $x,v,\partial/\partial x$ and $\partial/\partial v$ follows a simple pattern that repeats after two commutations. In particular, note that
\begin{align*}
&[x,Q]=b\frac{\partial}{\partial v}, &\left[ [x,Q],Q \right]&=abx, &\ldots \nonumber\\
&[\frac{\partial}{\partial x},Q]=av , &\left[ [\frac{\partial}{\partial x},Q],Q \right]&=ab\frac{\partial}{\partial x}, &\ldots \nonumber\\
&[v,Q]=b\frac{\partial}{\partial x}, &\left[ [v,Q],Q \right]&=abv, &\ldots\nonumber\\
&[\frac{\partial}{\partial v},Q]=ax, &\left[ [\frac{\partial}{\partial v},Q],Q \right]&=ab\frac{\partial}{\partial v},&\ldots
\end{align*}
Carrying out the nested commutators to all orders and summing the result yields, for $a,b>0$, the following 
\begin{align}
\label{eq:rep0}
e^{-\tau Q}xe^{\tau Q}&=\cosh(\tau \sqrt{ab})x+\sqrt{\frac{b}{a}}\sinh(\tau \sqrt{ab})\frac{\partial }{\partial v}\nonumber\\
e^{-\tau Q}\frac{\partial}{\partial x}e^{\tau Q}&=\cosh(\tau \sqrt{ab})\frac{\partial }{\partial x}+\sqrt{\frac{a}{b}}\sinh(\tau \sqrt{ab})v\nonumber\\
e^{-Q}ve^{\tau Q}&=\cosh(\tau \sqrt{ab})v+\sqrt{\frac{b}{a}}\sinh(\tau \sqrt{ab})\frac{\partial }{\partial x}\nonumber\\
e^{\tau -Q}\frac{\partial}{\partial v}e^{\tau Q}&=\cosh(\tau \sqrt{ab})\frac{\partial }{\partial v}+\sqrt{\frac{a}{b}}\sinh(\tau \sqrt{ab})x
\end{align}
The Euclidean result given above is simpler that the Minkowski commutators, which have alternating signs and $i$'s in the various expressions.

Consider the following equation
\begin{equation}
e^{-Q/2}He^{Q/2}=C_1\frac{\partial^2}{\partial v^2}+C_2x\frac{\partial}{\partial v}+C_3v\frac{\partial}{\partial x} +C_4\frac{\partial^2}{\partial x^2}+C_5x^2+C_6v^2
\label{eq:rep2}
\end{equation}

To obtain the factorization of the Hamiltonian into two de-coupled oscillators, choose the following values for $a$ and $b$, namely
\begin{equation}
\label{abqparamaet}
\sqrt{\frac{a}{b}}=\gamma\omega_1\omega_2;~~\sinh(\sqrt{ab})=\frac{2\omega_1\omega_2}{\omega_1^2-\omega_2^2}~~\Rightarrow~~\sqrt{ab}=\ln\left(\frac{\omega_1+\omega_2}{\omega_1-\omega_2}\right)
\end{equation}
Note the definition of  $a$ and $b$ is based on $\omega_1, \omega_2$ being real and $\omega_1>\omega_2$, and which makes the operator $Q$ Hermitian.

Define
\begin{align}
\label{coeffab}
A=\cosh(\frac{\sqrt{ab}}{2})&=\frac{\omega_1}{\sqrt{\omega_1^2-\omega_2^2}}~~;~~
B=\sinh(\frac{\sqrt{ab}}{2})=\frac{\omega_2}{\sqrt{\omega_1^2-\omega_2^2}}\\~
C&=\sqrt{\frac{a}{b}}=\gamma\omega_1\omega_2 \nonumber
\end{align}

Using the result of Eq.~\ref{eq:rep0} and the definitions in Eq.~\ref{eq:rep2} yields the following
\begin{align}
C_1&=-\frac{1}{2\gamma}A^2+\frac{\gamma}{2}\left(\frac{B}{C} \right)^2\omega_1^2 \omega_2^2\nonumber\\
&=-\frac{1}{2\gamma}\left[\cosh^2\left(\frac{\sqrt{ab}}{2}\right) -\sinh^2\left(\frac{\sqrt{ab}}{2}\right)\right]\nonumber\\
&=-\frac{1}{2\gamma}
\end{align}
Similarly, after some simplifications
\begin{align}
C_2&=-\frac{CAB}{\gamma}+\gamma\omega_1^2\omega_2^2\left(\frac{AB}{C}\right)=0\nonumber\\
C_3&=-(A^2+B^2)+\gamma(\omega_1^2+\omega_2^2)\left(\frac{AB}{C}\right)=0\nonumber
\end{align}
The constants $a$ and $b$ are chosen so that $C_2=C_3=0$; hence one has the following
\begin{equation}
C_2=C_3=0~~\Rightarrow~~~\text{Determines} ~a ~\text{and}~ b
\end{equation}
The remaining coefficients are given by
\begin{align}
C_4&=-\frac{AB}{C}+\frac{\gamma}{2}(\omega_1^2+\omega_2^2)\left( \frac{B}{C}\right)^2=-\frac{1}{2\gamma\omega_1^2}\nonumber\\
C_5&=-\frac{1}{2\gamma}B^2C^2+\frac{\gamma}{2}\omega_1^2\omega_2^2A^2=\frac{\gamma}{2}\omega_1^2\omega_2^2\nonumber\\
C_6&=-ABC+\frac{\gamma}{2}(\omega_1^2+\omega_2^2)A^2=\frac{\gamma}{2}\omega_1^2
\end{align}

Collecting all the results yields
\begin{align}
\label{eq:rep9}
e^{-Q/2}He^{Q/2}&=H_{O}\\
\label{eq:rep3}
&=-\frac{1}{2\gamma}\frac{\partial^2}{\partial v^2}-\frac{1}{2\gamma\omega_1^2}\frac{\partial^2}{\partial x^2}+\frac{\gamma}{2}\omega_1^2v^2+\frac{\gamma}{2}\omega_1^2 \omega_2^2x^2
\end{align}

\section{The matrix elements of $e^{-\tau Q}$} \label{sec:matrixqoper}
The $Q$-operator is given from Eq. \ref{Qoperdef} by
\begin{equation}
Q=axv-b\frac{\partial^2}{\partial x \partial v}
\end{equation}
where, from Eq. \ref{abqparamaet}
\begin{equation*}
\sqrt{\frac{a}{b}}=\gamma\omega_1\omega_2~;\quad \sinh(\sqrt{ab})=\frac{2\omega_1\omega_2}{\omega_1^2-\omega_2^2}
\end{equation*}

The finite matrix elements of the $e^{\pm\tau Q}$-operator are required for many calculations involving the state vectors. The exact matrix elements can be obtained by noting that  the Hermitian $Q$-operator factorizes into two decoupled harmonic oscillators by an appropriate change of variables.

Consider the change of variables given by
\begin{align}
\alpha&=\frac{1}{\sqrt 2}(x+v)~;\quad \beta=\frac{1}{\sqrt 2}(x-v)\\
 \frac{\partial}{\partial x}&=\frac{1}{\sqrt 2}(\frac{\partial}{\partial \alpha}+\frac{\partial}{\partial \beta})~;\quad  \frac{\partial}{\partial v}=\frac{1}{\sqrt 2}(\frac{\partial}{\partial \alpha}-\frac{\partial}{\partial \beta})
\end{align}
In these coordinates the $\alpha,~\beta$ sectors completely factorize and yield
\begin{align}
Q&=\frac{a}{2}(\alpha^2-\beta^2)-\frac{b}{2}(\frac{\partial^2}{\partial \alpha^2}-\frac{\partial^2}{\partial \beta^2}) \nonumber\\
\label{qfactor}
&=-\frac{1}{2}b\frac{\partial^2}{\partial \alpha^2}+\frac{1}{2}a\alpha^2+\frac{1}{2}b\frac{\partial^2}{\partial \beta^2}-\frac{1}{2}a\beta^2
\end{align}

Consider the Hamiltonian of a quantum oscillator given by
\begin{equation}
H_{\text{sho}}=-\frac{1}{2m}\frac{\partial^2}{\partial z^2}+\frac{1}{2}m\omega^2 z^2
\end{equation}
with the transition amplitude given by \cite{fey}
\begin{align}
&K(z,z';\tau)=\bra{z}e^{-\tau H}\ket{z'}\nonumber\\
\label{eq:K}
&=\sqrt{\frac{m\omega}{2\pi\sinh{\omega\tau}}} \exp\left\{-\frac{m\omega}{2\sinh(\omega\tau)}\Big[(z^2+z'^2)\cosh(\omega\tau)-2zz'\Big]\right\}
\end{align}
Comparing the $\alpha$- and $\beta$-sectors of $Q$ with $H_{\text{sho}}$ shows that, for $a, b$ real, the $\alpha$ sector is the usual quantum oscillator but the $\beta$ sector yields a divergent transition amplitude. The result for the $\beta$ sector is assumed to be given by the analytic continuation of the oscillator transition amplitude; this assumption will later be verified by an independent derivation.

To exploit the quadratic form of the $\alpha$ and $\beta$ sectors, consider extending the range of $m$ to the real line.
\begin{align}
&\alpha-\mathrm{sector}:  &b&=\frac{1}{m}; &a&=m\omega^2  & \Rightarrow& &\omega&=\sqrt{ab}\\
&\beta-\mathrm{sector}:  &b&=-\frac{1}{m}; &a&=-m\omega^2 & \Rightarrow& &\omega&=\sqrt{ab}
\label{eq:sector}
\end{align}

Hence Eqs. \ref{qfactor}, \ref{eq:K} and Eq.~\ref{eq:sector} yield the following
\begin{align*}
&\bra{\alpha,\beta}e^{-\tau Q}\ket{\alpha',\beta'}\\
&=N_\alpha N_\beta \exp \left\{-\frac{1}{2b} \frac{\sqrt{ab}}{ \sinh (\tau\sqrt{ab})}\big[(\alpha^2+\alpha'^2)\cosh(\tau\sqrt{ab})-2\alpha\alpha'\big]\right\}\\
&\qquad\times \exp \left\{\frac{1}{2b} \frac{\sqrt{ab}}{ \sinh(\tau\sqrt{ab})}\big[(\beta^2+\beta'^2)\cosh(\tau\sqrt{ab})-2\beta\beta'\big] \right\}\\
& =\mathcal{N}(\tau)\exp\Big\{-\frac{1}{2}\mathcal {G}(\tau)(\alpha^2+\alpha'^2-\beta^2-\beta'^2) + \mathcal {H}(\tau)(\alpha\alpha'-\beta\beta') \Big\}
\end{align*}
where
\begin{align}
&\mathcal {G}(\tau)=\sqrt{\frac{a}{b}}\coth(\tau\sqrt{ab})~;\quad \mathcal {H}(\tau)=\sqrt{\frac{a}{b}}\frac{1}{\sinh(\tau \sqrt{ab})}
\end{align}
and
\begin{align}
&\mathcal{N}(\tau)=\sqrt{\frac{\sqrt{ab}}{2\pi b \sinh(\tau\sqrt{ab})}}\cdot\sqrt{\frac{\sqrt{ab}}{-2\pi b \sinh(\tau\sqrt{ab})}}\nonumber\\
\label{coeffnorm}
&\Rightarrow \mathcal{N}(\tau) =\frac{i}{2\pi}\sqrt{\frac{a}{b}}\frac{1}{|\sinh(\tau\sqrt{ab})|}=\frac{i}{2\pi}\big |\mathcal {H}(\tau) \big|
\end{align}

Hence a heuristic derivation for $Q$ yields
\begin{align}
&\bra{x,v}e^{-\tau Q} \ket{x',v'}=\mathcal{N}(\tau)\exp\left\{-\mathcal {G}(\tau)(xv+x'v')+\mathcal {H}(\tau)(xv'+vx') \right\} 
\label{eq:AB}
\end{align}
The normalization constant $\mathcal{N}(\tau)$ will be seen to play a crucial role in the normalization of all the state vectors of the Hamiltonian $H$.

For the real branch, both coefficients  $\mathcal {G}(\tau)$ and $\mathcal {H}(\tau)$ are real and hence $Q=Q^\dagger$ is Hermitian. 

Note the form obtained for $e^{-\tau Q}$ in Eq.~\ref{eq:AB} is a major simplification since in general, for Hermitian $Q$, one would need to evaluate $4+12=16$ real coefficients. Instead, the form that has been heuristically derived in Eq.~\ref{eq:AB} has reduced the determination of $e^{-\tau Q}$ to that of computing two real coefficient functions $\mathcal {G}(\tau)$ and $\mathcal {H}(\tau)$ and a normalization constant $\mathcal N(\tau)$.

The operator $Q$ is unbounded and many of the manipulations are only formally valid. For example, the identity $e^{-\tau Q}e^{\tau Q}=\mathbb{ I}$ holds for the matrix elements only in a formal sense. Evaluating the matrix elements of the product  $e^{-\tau Q}e^{\tau Q}$  yields the following
\begin{align*}
&\bra{x,v}e^{-\tau Q} e^{\tau Q} \ket{x',v'}=\int d\xi d\zeta \bra{x,v}e^{-\tau Q}\ket{\xi,\zeta} \bra{\xi,\zeta}e^{\tau Q} \ket{x',v'}\\
&=\mathcal{N}^2(\tau)\int d\xi d\zeta e^{-\mathcal {G}(\tau)(xv+\xi\zeta)+\mathcal {H}(\tau)(x\zeta+v\xi) } e^{\mathcal {G}(\tau)(\xi\zeta+x'v')-\mathcal {H}(\tau)(\xi v'+\zeta x') } \\
&=\mathcal{N}^2(\tau) e^{-\mathcal {G}(\tau)(xv-x'v')} \int d\xi d\zeta \exp\{i\mathcal {H}(\tau) i\xi(v'-v)+i\mathcal {H}(\tau) i\zeta(x'-x)\}\\
&=\mathcal{N}^2(\tau)\left(\frac{2\pi}{i\mathcal {H}(\tau)}\right)^2 \delta(x-x')\delta(v-v')=\delta(x-x')\delta(v-v')
\end{align*}
Hence
\begin{align}
\label{eqeqone}
 e^{-\tau Q} e^{\tau Q} =\mathbb{I}
\end{align}
To make the above derivation more rigorous one can analytically continue $\tau$ back to Minkowski time $t=-i\tau$, do the computation and then analytically continue back to Euclidean time $\tau$. This would give the result given above.

\section{Vacuum state; $e^{\pm Q/2}$}
The vacuum state and the dual vacuum state of the Euclidean Hamiltonian is obtained using the expression for $e^{Q/2}$ and $e^{-Q/2}$ respectively. The vacuum can verified to be correct by the direct application of the Hamiltonians $H$ and $H^\dagger$ and hence provides an independent verification for the matrix elements obtained for $Q$ in Section \ref{sec:matrixqoper}.

From the oscillator Hamiltonian $H_0$ given in Eq.~\ref{eq:rep3}, the oscillator vacuum state, by inspection, is given by
\begin{eqnarray*}
&H_0\ket{0,0}=\frac{1}{2}(\omega_1 +\omega_2)\ket{0,0}\\
&\braket{x,v}{0,0}=\left(\frac{\gamma^2 \omega_1^3 \omega_2}{\pi^2} \right)^{1/4}\exp \left\{-\frac{\gamma}{2}(\omega_1v^2+\omega_1^2\omega_2x^2)\right\}
\end{eqnarray*}

The vacuum state is given by
\begin{align}
\label{vacuumhighenergy}
&\ket{\Psi_{00}}=e^{Q/2}\ket{0,0}~;~H\ket{\Psi_{00}}=E_0\ket{\Psi_{00}}~;~E_{00}=\frac{1}{2}[\omega_1+\omega_2]
\end{align}

The co-ordinate representation of the vacuum state can be directly obtained from the Hamiltonian $H$ and is given by \cite{mannheim1}
\begin{align}
\label{vacuumhigh}
&\Psi_{00}(x,v)=\langle x,v\ket{\Psi_{00}}=\bra{x,v}e^{Q/2}\ket{0,0}\nonumber\\
&~~~=N_{00}\exp\left\{-\frac{\gamma}{2}(\omega_1+\omega_2)\omega_1\omega_2x^2-\frac{\gamma}{2}(\omega_1+\omega_2)v^2-\gamma\omega_1\omega_2xv \right\}
\end{align}
The vacuum state $\ket{\Psi_{00}}$ is real valued and  normalizable, with $N_{00}$ the normalization constant.

It can be directly verified that the dual ground state, from Eq. \ref{vacuumhigh}, is given by
\begin{align*}
&\Psi_{00}^D(x,v)=N_{00}\exp\left\{-\frac{\gamma}{2}(\omega_1+\omega_2)\omega_1\omega_2x^2-\frac{\gamma}{2}(\omega_1+\omega_2)v^2+\gamma\omega_1\omega_2xv \right\}
\end{align*}

More formally, the dual vacuum state obeys the following
\begin{align}
&\bra{\Psi^D_{00}}=\bra{0,0}e^{-Q/2}~~~~~~~~~~~~\nonumber\\
\Rightarrow~~ &\Psi^D_{00}(x,v)=\bra{\Psi^D_{00}} x,v\rangle =\bra{0,0}e^{-Q/2}\ket{x,v}\nonumber
\end{align}

From Eq.~\ref{eq:AB}, $\tau=-1/2$ yields
\begin{align}
\bra{x,v}e^{Q/2}\ket{x',v'}= ~~~~~~~~~~~~~~~~~~~~~~~~~~~~~~\nonumber\\
\mathcal{N}(\frac{1}{2})\exp\left\{ \sqrt{\frac{a}{b}}\coth \left(\frac{\sqrt{ab}}{2} \right)(xv+x'v')-\sqrt{\frac{a}{b}} \frac{(xv'+vx')}{\sinh \left(\frac{\sqrt{ab}}{2}\right)} \right\} 
\label{qovertwo}
\end{align}
Note
\begin{align}
&\sqrt{\frac{a}{b}}\coth\left(\frac{\sqrt{ab}}{2} \right)=\gamma\omega_1^2~;\quad \sqrt{\frac{a}{b}}\frac{1}{\sinh \left(\frac{\sqrt{ab}}{2}\right) } =\gamma\omega_1\sqrt{\omega_1^2-\omega_2^2}\\
\label{coeffnhalf}
&\mathcal{N}(\frac{1}{2})
=\frac{i}{2\pi}\sqrt{\frac{a}{b}}\frac{1}{\sinh \left(\frac{\sqrt{ab}}{2}\right) }=\frac{i}{2\pi}\gamma\omega_1\sqrt{\omega_1^2-\omega_2^2}
\end{align}
Hence, from Eq. \ref{qovertwo}
\begin{align*}
&\bra{x,v}e^{Q/2}\ket{x',v'}=\mathcal{N}(\frac{1}{2})\exp \left\{\gamma\omega_1^2(xv+x'v')-\gamma\omega_1\sqrt{\omega_1^2-\omega_2^2}(xv'+vx') \right\}
\end{align*}
Hence
\begin{eqnarray*}
\Psi_{00}(x,v)&=&\bra{x,v}e^{Q/2}\ket{0,0}\\
&=&\int d\xi d\zeta \bra{x,v}e^{Q/2}\ket{\xi,\zeta}\braket{\xi,\zeta}{0,0}\\
&=&\mathcal{N}(\frac{1}{2}) \left(\frac{\gamma^2 \omega_1^3 \omega_2}{\pi} \right)^{1/4}\int d\xi d\zeta e^S
\end{eqnarray*}
where
\begin{align*}
&S=-\frac{\gamma}{2}(\omega_1\zeta^2+\omega_1^2\omega_2\xi^2-2c_1\xi\zeta)-\gamma c_2(\zeta x+v\xi)+\gamma c_1xv\\
&c_1=\omega_1^2~;\quad c_2 =\omega_1\sqrt{\omega_1^2-\omega_2^2}
\end{align*}
Performing the Gaussian integrations over $\xi$ and $\zeta$ yields 
\begin{equation*}
\Psi_{00}(x,v)=N_{00}e^F
\end{equation*}
where
\begin{align*}
F&=\frac{\gamma^2 c_2^2	}{2\gamma(\omega_1^3\omega_2-c_1^2)}[\omega_1v^2+\omega_1^2\omega_2x^2+2c_1xv]+\gamma c_1xv\\
&=-\frac{\gamma}{2}[(\omega_1+\omega_2)\omega_1\omega_2x^2+(\omega_1+\omega_2)v^2]-\gamma\omega_1\omega_2xv
\end{align*}
which is the expected result.

To determine $N_{00}$ note that the Gaussian integration has the following matrix
\begin{eqnarray*}
 &&M=\gamma\left[ 
\begin{array}{ll}   
\omega_1^2\omega_2 & -c_1 \nonumber\\
-c_1& \omega_1  \end{array} \right] ~~;~~\sqrt{\det(M)}=i\gamma\omega_1^{3/2} \sqrt{\omega_1-\omega_2}
\end{eqnarray*}
Note the determinant of matrix $M$ is \textit{negative}  and hence leads to the correct sign for the exponent $F$ of the vacuum state vector.

Performing the Gaussian integration and using Eq. \ref{coeffnhalf} yields the following
\begin{eqnarray}
&&N_{00}= \mathcal{N}(\frac{1}{2}) \left(\frac{\gamma^2 \omega_1^3 \omega_2}{\pi} \right)^{1/4} \frac{2\pi}{\sqrt{\det(M)}}\nonumber\\
&&=\frac{i}{2\pi}\gamma\omega_1\sqrt{\omega_1^2-\omega_2^2} \left(\frac{\gamma^2 \omega_1^3 \omega_2}{\pi^2} \right)^{1/4} \frac{2\pi}{i\gamma\omega_1^{3/2} \sqrt{\omega_1-\omega_2}} \nonumber\\
\label{normzerozero}
&&\Rightarrow N_{00}=(\omega_1 \omega_2)^{1/4} \sqrt{\frac{\gamma}{\pi}(\omega_1+\omega_2)}
\end{eqnarray}

To derive the dual vacuum state, note from Eq. \ref{qovertwo}
\begin{align*}
&\bra{x,v}e^{-Q/2}\ket{x',v'}=\mathcal{N}(\frac{1}{2})\exp \left\{-\gamma\omega_1^2(xv+x'v')+\gamma\omega_1\sqrt{\omega_1^2-\omega_2^2}(xv'+vx') \right\}
\end{align*}
A derivation similar to the one for the vacuum state $\Psi_{00}$ yields the dual vacuum state
\begin{align}
\label{vacqderivdual} 
\Psi_{00}^D(x,v)&=\bra{0,0}e^{-Q/2}\ket{x,v}\nonumber\\
&=N_{00}\exp\{-\frac{\gamma}{2}(\omega_1+\omega_2)\omega_1\omega_2x^2-\frac{\gamma}{2}(\omega_1+\omega_2)v^2+\gamma\omega_1\omega_2xv\}
\end{align}

The norm of a state is defined by the scalar product of state with its dual and yields the following norm for the vacuum state
\begin{align*}
\langle \Psi_{00}^D|\Psi_{00}\rangle &=\int dxdv\Psi_{00}^D(x,v)\Psi_{00}(x,v)\\
&=N_{00}^2\int dxdv \exp\{-\gamma(\omega_1+\omega_2)\omega_1\omega_2x^2-\gamma(\omega_1+\omega_2)v^2\}\\
&=1
\end{align*}

\section{Heisenberg operator equations}
In Schr\"odinger's formulation of quantum mechanics, all the time dependence of a quantum system arises due to the time evolution of the  state vector $|\Psi(t_M)\rangle$, where $t_M$ is Minkowski time;  the operators $\mathcal{O}$ being taken to be time-independent. The Schr\"odinger equation yields the following
\begin{align}
&|\Psi(t_M)\rangle=e^{-it_MH}|\Psi\rangle~~;~~\langle \chi(t_M)|=\langle \chi|~e^{it_MH}
\end{align}
with the time-dependent expectation of operator $\mathcal{O}$ given by
\begin{align}
&E[\mathcal{O};t_M]=\langle \chi(t_M)|\mathcal{O}|\Psi(t_M)\rangle=\langle \chi|e^{it_MH}\mathcal{O} e^{-it_MH}|\Psi\rangle
\end{align}

In Heisenberg's formulation of quantum mechanics, all the time dependence of an operator $\mathcal{O}_H(t_M)$ arises due to the time evolution of the operators, with the state vector  $|\Psi\rangle$ and the dual state vectors $\langle \chi|$ being taken to be time-independent; the expectation value is then given by 
\begin{align}
&E[\mathcal{O};t_M]=\langle \chi|\mathcal{O}_H(t_M)|\Psi\rangle\\
&\mathcal{O}_H(t)=e^{it_MH}\mathcal{O} e^{-it_MH}
\end{align}

For Euclidean time $\tau$, one has the following expressions
\begin{align*}
&\tau=-it_M\\
&\mathcal{O}_H(\tau)=e^{\tau H}\mathcal{O} e^{-\tau H}
\end{align*}
and yields Heisenberg operator equations of motion for Euclidean time 
\begin{align*}
&\frac{\partial \mathcal{O}_H(\tau)}{\partial \tau}=[H,\mathcal{O}_H(\tau)]
\end{align*}

The non-Hermitian Hamiltonian $H$ for action with acceleration, from Eq. \ref{eq:rep9}, is given by
\begin{align}
 H^\dagger&=e^{-Q/2}H_{O} e^{Q/2} \\
\Rightarrow & H^\dagger=e^{-Q}He^{Q}:~~\text{pseudo-Hermitian}\\
\label{exphhdag}
\Rightarrow &\left(e^{-\tau H}\right)^\dagger=e^{-\tau H^\dagger}=e^{-Q}e^{-\tau H}e^{Q}
\end{align}

To obtain the Heisenberg operator equations of motion, it is necessary to define the dual state vectors for the pseudo-Hermitian Hamiltonian using $Q$-conjugation. Namely
\begin{align}
&\ket{\chi}~~\to~~\langle \chi|e^{-Q}
\end{align}

Hence, for Euclidean time, the time dependent expectation value, using Eq. \ref{exphhdag} yields the following result
\begin{align}
&E[\mathcal{O};\tau]=\langle \chi(\tau)|e^{-Q}\mathcal{O}|\Psi(\tau)\rangle=\langle \chi|e^{\tau H^\dagger}e^{-Q} \mathcal{O} e^{-\tau H}|\Psi\rangle\nonumber\\
\label{hesispeduo}
&=\langle \chi|e^{-Q} e^{\tau H}\mathcal{O} e^{-\tau H}|\Psi\rangle=\langle \chi|e^{-Q} \mathcal{O}_H(\tau)|\Psi\rangle
\end{align}

Hence, from Eq. \ref{hesispeduo}, due to the choice of the Hilbert space metric $e^{-Q}$, the Heisenberg time-dependent operator $\mathcal{O}_H(\tau)$ for the pseudo-Hermitian Hamiltonian is given by the same expression as for the Hermitian Hamiltonian, namely
\begin{align*}
&\mathcal{O}_H(\tau)=e^{\tau H}\mathcal{O} e^{-\tau H}\\
&\Rightarrow \frac{\partial \mathcal{O}_H(\tau)}{\partial \tau}=[H,\mathcal{O}_H(\tau)]
\end{align*}
In particular, for the acceleration  Hamiltonian  given by Eq. \ref{eq:Hamiltonian1}
\begin{equation*}
H=-\frac{1}{2\gamma}\frac{\partial^2}{\partial v^2}-v\frac{\partial}{\partial x}+\frac{1}{2}\alpha v^2+\Phi(x)
\end{equation*}
the Heisenberg equation for the position operator $x_H(\tau)$, with $x$ being the Schr\"odinger position operator, yields the following
\begin{align}
&{x}_H(\tau)=e^{\tau H}x e^{-\tau H}\nonumber\\
\label{vequaltoxdot}
& \dot{x}_H(\tau)=\frac{\partial {x}_H(\tau)}{\partial \tau}=[H,\mathcal{O}_H(\tau)]=-v_H(\tau)
\end{align}
Hence, the identification made in the path integral derivation, namely  $\dot{x}(\tau)=-v(\tau)$, is seen to also hold as an operator equation for the Heisenberg operators $\dot{x}_H(\tau), v_H(\tau)$ as shown by Eq. \ref{vequaltoxdot}.

To illustrate the role of the Hilbert space metric $e^{-Q}$ consider the time-ordered vacuum expectation value of the Heisenberg operators at two different (Euclidean) times $\tau>0$. The vacuum state is given by $|\Psi_{00}\rangle$ with $H\Psi_{00}\rangle=E_{00}|\Psi_{00}\rangle;~E_{00}\equiv E_0=(\omega_1+\omega_2)/2$; using the rule for forming the dual vector of the pseudo-Hermitian Hamiltonian yields the following
\begin{align}
& G(\tau)=\langle \Psi^D_{00}|{x}_H(\tau){x}_H(0) |\Psi_{00}\rangle=\langle \Psi_{00}|e^{-Q}{x}_H(\tau){x}_H(0) |\Psi_{00}\rangle  \nonumber\\
\label{proeqnvev}
&~~~=\langle \Psi_{00}|e^{-Q}e^{\tau H}x e^{-\tau H}x |\Psi_{00}\rangle=\langle \Psi_{00}|e^{\tau H^\dagger}e^{-Q}x e^{-\tau H}x |\Psi_{00}\rangle\nonumber\\
&~~~~~~~=\langle \Psi_{00}|e^{-Q}x e^{-\tau (H-E_0)}x |\Psi_{00}\rangle
\end{align}
since $\langle \Psi^D_{00}|H^\dagger=E_0\langle \Psi^D_{00}|$.

The propagator $G(\tau)$ in Eq. \ref{proeqnvev} is the propagator and is analyzed in detail in \cite{bebeujordan}.

Recall that the probability amplitude is given in Eq. \ref{kerstasp}
\begin{equation*}
\mathcal{K}_S(x,v;x',v')=\bra{x,v}e^{-\tau H}\ket{x',v'}
\end{equation*}
and the matrix elements of the the Hilbert space metric $e^{-Q}$ is given (later) in Eq. \ref{eq:AB}
\begin{align}
&\bra{x,v}e^{-\tau Q} \ket{x',v'}=N\exp\left\{-\mathcal {G}(\tau)(xv+x'v')+\mathcal {H}(\tau)(xv'+vx') \right\} 
\end{align}
For both the operators $e^{-\tau H}$ and $e^{-\tau Q}$, there is no need for an extra metric $e^{-\tau Q}$ since Eqs. \ref{kerstasp} and \ref{eq:AB} are the matrix elements of the operators in a complete basis $\langle x,v|$ and its dual $|x',v'\rangle$. 

It is only the matrix elements of operators $e^{-\tau H}$ and so on  are determined for \textit{eigenstates} and \textit{dual eigenstates} of $H$, then the metric $e^{- Q}$ is required.

A verification of the definition of $\mathcal{K}_S(x,v;x',v')$ is given by considering the limit of the probability amplitude for $\tau \to \infty$; Eq. \ref{kerstasp}  yields ($E_1$ is the generic energy of the state above the ground state)
\begin{eqnarray}
\lim_{\tau \to \infty}\mathcal{K}_S(x,v;x',v')\simeq e^{-\tau E_{00}} \bra{x,v}e^{Q/2}|0,0\rangle \langle 0,0|e^{-Q/2}\ket{x',v'} +O(e^{-(E_1-E_{00})\tau})\nonumber\\
\label{kerlargetau}
=e^{-\tau E_{00}}\Psi_{00}(x,v)\Psi_{00}^D(x',v')=e^{-\tau E_{00}}\Psi_{00}(x,v)\Psi_{00}(x',-v')~~~~~~~~~~~
\end{eqnarray}
In Appendix B, the value of $\mathcal{K}_S(x,v;x',v')$ for large $\tau$ is obtained from the classical solution in Eq. \ref{vacaction} and confirms the form of the kernel given in Eq. \ref{kerlargetau} is correct and consistent with its path integral definition.

One can evaluate the matrix elements of $e^{-\tau H}$ in Hilbert space using the basis states $e^{Q/2}\ket{x,v}$, the dual basis $\big(\bra{x,v}e^{Q/2}\big)e^{-Q}=\bra{x,v}e^{-Q/2}$ and the metric  $e^{-Q}$, and yields the following
\begin{eqnarray}
\label{matrixhzero}
&&\bra{x,v}e^{-Q/2}e^{-\tau H} e^{Q/2}\ket{x',v'}= \bra{x,v}e^{-\tau H_0} \ket{x',v'} 
\end{eqnarray}
The symmetry of the matrix elements of $\bra{x,v}e^{-Q/2}e^{-\tau H} e^{Q/2}\ket{x',v'}$ given  above in Eq. \ref{matrixhzero}  does not yield the acceleration Lagrangian that results for  $\bra{x,v}e^{-\tau H} \ket{x',v'}$;  hence the correct expression for the kernel is given by $\mathcal{K}_S(x,v;x',v')=\bra{x,v}e^{-\tau H} \ket{x',v'}$, and which in turn yields the correct matrix elements given by the path integral.

\section{Complex $\omega_1,\omega_2$}
An interesting case for the parameters is the complex domain for which $\omega_1,\omega_2$ are complex; the eigenfunctions and eigenenergies are well behaved as long as the vacuum state $\ket{\Psi_{00}}$ given in Eq. \ref{vacuumhigh} is normalizable. Let
\begin{align}
\omega_1=Re^{i\phi}~~;~~\omega_2=Re^{-i\phi}\\
\Rightarrow~~\omega_1+\omega_2=2R\cos(\phi)~~;~~\omega_1\omega_2=R^2
\end{align}
that yields, from Eq. \ref{vacuumhigh}, the vacuum state for the complex domain as
\begin{align}
\label{vacuumhighcom}
\Psi_{00}(x,v)&=N_{00}\exp\left\{-\gamma R^3\cos(\phi)x^2-\gamma R\cos(\phi)v^2-\gamma R^2xv \right\}
\end{align}
Hence, from Eq. \ref{vacuumhighcom}, the vacuum state and Hilbert space for the complex domain is well defined for
\begin{align}
\cos(\phi)>0~~\Rightarrow~~-\pi/2<\phi< \pi/2\nonumber
\end{align}

\section{Many degrees of freedom}
Consider the Hamiltonian that is a quadratic generalization of the acceleration action given in Eq. \ref{eq:Hamiltonian}. For degrees of freedom $x_n,~n=1,2,...N$ the acceleration  Hamiltonian , in matrix notation, is given as follows
\begin{equation}
\label{eq:Hamiltonianmany}
H=-\frac{1}{2}\frac{\partial}{\partial v}S^T\frac{1}{\gamma}S\frac{\partial}{\partial v}- v \frac{\partial}{\partial x}+\frac{1}{2}vS^T \alpha Sv+\frac{1}{2}xS^T\beta Sx
\end{equation}
A specific choice is made for the Hamiltonian that is simple and consistent with the definitions of $v$ and $x$ given in the defining Eq. \ref{cnstrxv}.
The Lagrangian is given by
\begin{equation}
\label{lagpotmany2}
\mathcal{L}=-\frac{1}{2}\left[\ddot x S^T\gamma S \ddot x+\dot xS^T \alpha S\dot x +xS^T \beta S x \right]
\end{equation}
The (real) orthogonal matrix $S$ and the diagonal matrices are given in matrix notation as follow
\begin{eqnarray}
\label{dmatridiag}
&&SS^T=\mathbb{I}~~;~~\gamma=\text{diag}(\gamma_1,\gamma_2,..,\gamma_N)\nonumber\\
&&\alpha=\text{diag}(\alpha_1,\alpha_2,..,\alpha_N)~~;~~\beta=\text{diag}(\beta_1,\beta_2,..,\beta_N)
\end{eqnarray}
Define new variables, in matrix notation
\begin{eqnarray}
\label{newvarwz}
x=S^T z;~~v=S^T u
\end{eqnarray}
The Hamiltonian in Eq. \ref{eq:Hamiltonianmany} is given by
\begin{equation}
H=-\frac{1}{2}\sum_{n=1}^N\frac{1}{\gamma_n}\frac{\partial^2}{\partial u_n^2}-\sum_{n=1}^N u_n\frac{\partial}{\partial z_n}+\frac{1}{2} \sum_{n=1}^N \alpha_nu_n^2+\frac{1}{2}\sum_{n=1}^N\beta_nz_m^2
\end{equation}
The corresponding Lagrangian, from Eq. \ref{lagpotmany2} and similar to Eq. \ref{lagpot},   is given by
\begin{equation}
\label{lagpotmany}
\mathcal{L}=-\frac{1}{2}\sum_{n=1}^N\gamma_n\left[\ddot z_n^2+(\omega_{1n}^2+\omega_{2n}^2)\dot z_n^2+\omega_{1n}^2\omega_{2n}^2z_n^2 \right]
\end{equation}
The parametrization is a generalization of Eq. \ref{omega1omega2} and is given by
\begin{eqnarray}
\label{omega1omega2many}
&&\omega_{1n}=\frac{1}{2\sqrt{\gamma_n}}\left(\sqrt{\alpha_n+2\sqrt{\gamma_n\beta_n}}+\sqrt{\alpha_n-2\sqrt{\gamma_n\beta_n}}\right)\\
&&\omega_{2n}=\frac{1}{2\sqrt{\gamma_n}}\left(\sqrt{\alpha_n+2\sqrt{\gamma_n\beta_n}}-\sqrt{\alpha_n-2 \sqrt{\gamma_n\beta_n}}\right)\\
&&\omega_{1n}>\omega_{2n}~~\text{for $\omega_{1n},~\omega_{2n}$ real} \nonumber
\end{eqnarray}
The diagonal $H$ given in Eq. \ref{eq:Hamiltonianmany} is obtained from a $Q$-operator, which is a generalization of Eq. \ref{Qoperdef}, and is given by
\begin{align}
\label{Qoperdefmany}
Q&=\sum_{n=1}^N Q_n\\
Q_n&=a_nz_nu_n-b_n\frac{\partial^2}{\partial z_n \partial u_n}
\end{align}
with the following values for $a_n$ and $b_n$
\begin{equation}
\sqrt{\frac{a_n}{b_n}}=\gamma_n\omega_{1n}\omega_{2n};~~\sqrt{a_nb_n}=\ln\left(\frac{\omega_{1n}+\omega_{2n}}{\omega_{1n}-\omega_{2n}}\right)
\end{equation}
The diagonal Hamiltonian $H_O$ is given similar to the earlier case. The ground state is given by generalizing Eq. \ref{vacuumhigh} and yields the  following
\begin{eqnarray}
\Psi_{00}(x,v)=
\mathcal{N}\prod_{n=1}^N \exp\left\{-\frac{\gamma_n}{2}(\omega_{1n}+\omega_{2n})\omega_{1n}\omega_{2n}z_n^2-\frac{\gamma_n}{2}(\omega_{1n}+\omega_{2n})u_n^2-\gamma_n\omega_{1n}\omega_{2n} z_nu_n \right\}\nonumber~~~
\end{eqnarray}
In terms of the original coordinates
\begin{eqnarray}
\label{vacuumhighmany}
\Psi_{00}(x,v)
=\mathcal{N} \exp\left\{-\frac{1}{2}\sum_{m,n=1}^N\Big(P_{mn} x_mx_n+Q_{mn}v_m v_n+2R_{mn} x_nv_n\Big) \right\}~~~
\end{eqnarray}
where, again in matrix notation
\begin{eqnarray*}
&&P=SpS^T;~~p_n=\gamma_n(\omega_{1n}+\omega_{2n})\omega_{1n}\omega_{2n}\\
&&Q=SqS^T;~~q_n=\gamma_n(\omega_{1n}+\omega_{2n})\\
&&R=SrS^T;~~r_n=\gamma_n\omega_{1n}\omega_{2n}\\
\end{eqnarray*}
\section{Conclusions}
The Euclidean acceleration action yields a well defined quantum system that has a well defined Euclidean Hamiltonian. The Euclidean path integral is the appropriate formulation for numerical and Monte Carlo simulation of the system.  The path integral using the action and the Hamiltonian yield the same result due to a constraint term in the Hamiltonian.  

The state space of the pseudo-Hermitian Euclidean Hamiltonian has a state space metric $e^{-Q}$, which is a natural generalization of the state space of quantum mechanics.  The metric $e^{-Q}$ is an unbounded operator that maps the non-Hermitian Hamiltonian to the oscillator Hamiltonian; the matrix elements of $e^{-Q}$ were evaluated exactly and shown to yield the expected similarity transformation.

As was seen in various derivations carried out, the results from Minkowski time serve as a useful guideline for the derivations, but given a plethora of $i$ and various $\pm$ signs that differ between the Euclidean and Minkowski results, all the derivations for the Euclidean have to done independently from the Minkowski case, and from first principles. 

\section{Acknowledgment}
I thank Cao Yang for useful discussions and Wang Qinghai for discussing and sharing many of his valuable insights.

\appendix

\section{$e^{-\tau Q}$ and similarity transformations}
The heuristic derivation for $\mathcal {G}(\tau)$ and $\mathcal {H}(\tau)$  was obtained by an analogy with the oscillator Hamiltonian and cannot be assumed to be correct since the $\beta$-sector yields an un-stable Hamiltonian. The result needs to be independently verify. 

The fundamental similarity transformations by $e^{\pm\tau Q}$ of operators $x,\partial /\partial x, v$ and $\partial /\partial v$ are directly obtained using the result given in Eq.~\ref{eq:AB} and shown to be identical to the  defining equations for $Q$ given in Eq.~\ref{eq:rep0}.

Recall from Eq. \ref{eq:rep0} that $e^{\pm\tau Q}$  yields the following similarity transformations
\begin{align*}
&I.~e^{- \tau Q}xe^{ \tau Q}=\cosh(\tau\sqrt{ab})x+\sqrt{\frac{b}{a}}\sinh(\tau\sqrt{ab})\frac{\partial }{\partial v}\nonumber\\
&II.~e^{- \tau Q}\frac{\partial}{\partial x}e^{ \tau Q}=\cosh(\tau\sqrt{ab})\frac{\partial }{\partial x}+\sqrt{\frac{a}{b}}\sinh(\tau\sqrt{ab})v\nonumber\\
&III.~e^{- \tau Q}ve^{ \tau Q}=\cosh(\tau\sqrt{ab})v+\sqrt{\frac{b}{a}}\sinh(\tau\sqrt{ab})\frac{\partial }{\partial x}\nonumber\\
&IV.~e^{- \tau Q}\frac{\partial}{\partial v}e^{ \tau Q}=\cosh(\tau\sqrt{ab})\frac{\partial }{\partial v}+\sqrt{\frac{a}{b}}\sinh(\tau\sqrt{ab})x
\end{align*}

Consider the operator equation
\begin{equation}
e^{- \tau Q}xe^{ \tau Q}=\cosh(\tau\sqrt{ab})x+\sqrt{\frac{b}{a}}\sinh(\tau\sqrt{ab})\frac{\partial }{\partial v}
\end{equation}
The matrix element of $e^{- \tau Q}xe^{ \tau Q}$, using $I$ from above, is given by
\begin{align}
\label{eq:qxq}
\bra{x,v}e^{- \tau Q}xe^{ \tau Q}\ket{x',v'}&=\cosh(\tau\sqrt{ab})x\delta(x-x')\delta(v-v') \nonumber\\
&\quad +\sqrt{\frac{b}{a}}\sinh(\tau\sqrt{ab})\frac{\partial }{\partial v}\delta(x-x')\delta(v-v')
\end{align}

Eq.~\ref{eq:AB} yields
\begin{equation}
\label{eminusQ}
\bra{x,v}e^{- \tau Q}\ket{x',v'}=N\exp\{-g(xv+x'v')+h(xv'+vx')\}
\end{equation}
where
\begin{eqnarray*}
 &&g=\mathcal {G}(\tau)=\sqrt{\frac{a}{b}}\frac{\cosh(\tau\sqrt{ab})}{\sinh(\tau\sqrt{ab})}~;~ h=\mathcal {H}(\tau)=\sqrt{\frac{a}{b}}\frac{1}{\sinh(\tau\sqrt{ab})}\\
&&N=\mathcal{N}(\tau)=\frac{i}{2\pi \mathcal {H}(\tau)}
\end{eqnarray*}
The left hand side of Eq.~\ref{eq:qxq} yields
\begin{align*}
\bra{x,v}e^{- \tau Q}xe^{ \tau Q}\ket{x,v}&=\int d\xi d\zeta \bra{x,v} e^{- \tau Q}\ket{\xi,\zeta}\bra{\xi,\zeta}\xi e^{ \tau Q} \ket{x',v'}\\
&=N\int d\xi d\zeta e^{-g(xv+\xi\zeta)+h(x\zeta+v\xi)}~\xi \bra{\xi,\zeta}e^{ \tau Q} \ket{x',v'}
\end{align*}
But
\begin{align}
&\xi \exp\{-g(xv+\xi\zeta)+h(x\zeta+v\xi)\} \nonumber \\
&= \left[\sqrt{\frac{b}{a}}\sin(\tau\sqrt{ab})\frac{\partial}{\partial v}+\cosh(\tau\sqrt{ab})x \right]e^{-g(xv+\xi\zeta)+h(x\zeta+v\xi)}
\label{eq:xiAB}
\end{align}
Hence, using $e^{- \tau Q}e^{ \tau Q}=\mathbb{I}$ given in Eq. \ref{eqeqone}, we obtain the expected result that
\begin{eqnarray*}
&\bra{x,v}e^{- \tau Q}xe^{ \tau Q}\ket{x',v'}=\left[\cosh(\tau\sqrt{ab})+\sqrt{\frac{b}{a}}\sinh(\tau\sqrt{ab})\frac{\partial}{\partial v} \right]\bra{x,v} e^{- \tau Q}e^{ \tau Q}\ket{x',v'}\\
&\Rightarrow I.~e^{- \tau Q}x e^{ \tau Q} =\cosh(\tau\sqrt{ab})+\sqrt{\frac{b}{a}}\sinh(\tau\sqrt{ab})\frac{\partial}{\partial v} 
\end{eqnarray*}

Consider
\begin{align}
&\bra{x,v}e^{- \tau Q}ve^{ \tau Q}\ket{x',v'}=\int d\xi d\zeta~ \bra{x,v}e^{- \tau Q}\ket{\xi,\zeta}\bra{\xi,\zeta}\zeta e^{ \tau Q}\ket{x',v'} \nonumber\\
&=\int d\xi d\zeta~ \zeta \bra{x,v}e^{- \tau Q}\ket{\xi,\zeta}\bra{\xi,\zeta}e^{ \tau Q}\ket{x',v'}\nonumber\\
\label{zetarep}
&=\left(\sqrt{\frac{b}{a}}\sinh(\tau\sqrt{ab})\frac{\partial}{\partial x}+\cosh(\tau\sqrt{ab})v \right)\delta(x-x')\delta(v-v')
\end{align}
and yields the expected result that
\begin{equation*}
II.~e^{- \tau Q}ve^{ \tau Q}=\cosh(\tau\sqrt{ab})v+\sqrt{\frac{b}{a}}\sinh(\tau\sqrt{ab})\frac{\partial}{\partial x}
\end{equation*}

Consider the following  matrix element 
\begin{align*}
&\bra{x,v}e^{- \tau Q}\frac{\partial}{\partial x}e^{ \tau Q}\ket{x',v'} =\int d\xi d\zeta \bra{x,v}e^{- \tau Q}\ket{\xi,\zeta}\bra{\xi,\zeta}\frac{\partial}{\partial \xi} e^{ \tau Q} \ket{x',v'}\\
&\quad =N\int d\xi d\zeta \left[ \left(- \frac{\partial}{\partial \zeta}\right)e^{-g(xv+\xi\zeta)+h(x\zeta+v\xi)} \right]\bra{\xi,\zeta} e^{ \tau Q} \ket{x',v'}\\
&\quad =\int d\xi d\zeta \Big[g\zeta-h v\Big]\bra{x,v}e^{- \tau Q}\ket{\xi,\zeta}\bra{\xi,\zeta} e^{ \tau Q} \ket{x,v}
\end{align*}

Using Eq.~\ref{zetarep} to replace $\zeta$ in above expression yields,
\begin{align}
e^{- \tau Q}\frac{\partial}{\partial x}e^{ \tau Q}&= g \left(\sqrt{\frac{b}{a}}\sinh(\tau\sqrt{ab})\frac{\partial }{\partial v}+\cosh(\tau\sqrt{ab})v\right) -hv  \nonumber\\
\Rightarrow III.~e^{- \tau Q}\frac{\partial}{\partial x}e^{ \tau Q}&=\cosh(\tau\sqrt{ab})\frac{\partial}{\partial x}+\sqrt{\frac{b}{a}}\sinh(\tau\sqrt{ab})v \nonumber
\end{align}

And lastly, similar to above derivation
\begin{align*}
&\bra{x,v}e^{- \tau Q}\frac{\partial}{\partial v}e^{ \tau Q}\ket{x',v'}=N\int d\xi d\zeta \left[ - \frac{\partial}{\partial \zeta} e^{-g(xv+\xi\zeta)+h(x\zeta+v\xi)} \right] \bra{\xi,\zeta} e^{ \tau Q} \ket{x',v'}\\
&=\int d\xi d\zeta \Big[g\xi-h x\Big]  \bra{x,v}e^{- \tau Q}\ket{\xi,\zeta}\bra{\xi,\zeta}e^{ \tau Q}\ket{x',v'}
\end{align*}

Using Eq.~\ref{eq:xiAB} to replace $\xi$ in equation above yields
\begin{align*}
e^{- \tau Q}\frac{\partial}{\partial v}e^{ \tau Q}&=g\left(\sqrt{\frac{b}{a}}\sinh(\tau\sqrt{ab})\frac{\partial }{\partial v}+ \cosh(\tau\sqrt{ab})x \right)-hx\\
\Rightarrow IV.~ e^{- \tau Q}\frac{\partial}{\partial v}e^{ \tau Q}&=\cosh(\tau\sqrt{ab})\frac{\partial}{\partial v}+\sqrt{\frac{b}{a}}\cosh(\tau\sqrt{ab})x
\end{align*}

The finite matrix elements of $e^{- \tau Q}$ produce all the defining similarity transformations on the $x,v$ degrees of freedom and verifies that the heuristic derivation of the matrix elements of $e^{- \tau Q}$ is, in fact, correct.

\section{The classical solution} 
Consider the following parametrization of the acceleration Lagrangian
\begin{equation}
\label{clsolnpara}
 \mathcal L=-\frac{1}{2} \Big(a \ddot x^2+2 b(\dot x)^2+c  x^2 \Big) ~~;~~S=\int_0^\tau dt  \mathcal L
 \end{equation}
The parametrization chosen in Eq. \ref{clsolnpara} is more suitable for studying the classical solutions and is different from the one given in  Eq. \ref{lagpot}.
 
The Euler-Lagrangian equation
\begin{equation}
\left.\frac{\delta S}{\delta x} \right|_{x=x_c}=0 
\label{eq:eqofmotionbs} 
 \end{equation}
yields the equation of motion; the classical solution $x_c(t)$ satisfies following equation of motion
\begin{equation}
a \ddddot x_c(t) - 2 b  \ddot x_c(t) + c x_c(t) = 0.
\label{eq:eqofmotion}
 \end{equation}
 
Choose the  boundary conditions
\begin{align}
\text{Initial values}~~:~~x(0)&=x_f=x_1, \quad \dot x(0)=\dot x_f=x_2=-v_f \nonumber\\
\text{Final values}~~:~~x(\tau) &= x_i=x_4, \quad \dot x(\tau)=\dot x_i=x_3=-v_i
\label{eq:BC2}
\end{align}

From the definition of the probability amplitude given in Eq. \ref{kerstasp}
\begin{equation*}
\mathcal{K}_S(x_f,v_f;x_i,v_i)=\bra{x_f,v_f}e^{-\tau H}\ket{x_i,v_i}=\int DX e^S
\end{equation*}
with boundary conditions given by Eq. \ref{eq:BC2}.

Define the parameters  $r$ and $\omega$ by
\begin{equation}
 r + i \omega   \equiv \sqrt{\frac{b + i \sqrt{a c  -b^2}}{ a}} 
\end{equation}

From Eq.~\ref{eq:eqofmotion} the classical solution of equations of motion is given by \cite{cythesis}
\begin{equation}
x_c(t)=e^{rt}(a_1 \sin \omega t+a_2 \cos \omega t)+e^{-rt}(a_3 \sin \omega t+a_4 \cos \omega t)
\end{equation}
The parameters $a_1,..,a_4$ are obtained from the boundary conditions and given by the following.
\begin{align*}
a_1&=\Gamma\Bigl[ r^2 x_f e^{2 r \tau } \sin (2 \tau  \omega )+\omega  v_f e^{2 r \tau }-r v_f e^{2 r \tau } \sin (2 \tau  \omega )+r \omega  x_f e^{2 r \tau } \cos (2 \tau  \omega
   )-r \omega  x_f\nonumber\\
   &\qquad-\omega  v_f
    -2 r^2 x_i e^{r \tau } \sin (\tau  \omega )-2 r v_i e^{r \tau } \sin (\tau  \omega )-\omega  e^{r \tau } \left(e^{2 r \tau }-1\right)
   \cos (\tau  \omega ) \left(v_i+r x_i\right)\nonumber\\
   &\qquad-\omega ^2 x_i e^{r \tau } \sin (\tau  \omega )+\omega ^2 x_i e^{3 r \tau } \sin (\tau  \omega )\Bigl]
\end{align*}
\begin{align*}
a_2&=\Gamma\Bigl[r^2 x_f \left(-e^{2 r \tau }\right)+r v_f e^{2 r \tau }+r e^{2 r \tau } \cos (2 \tau  \omega ) \left(r x_f-v_f\right)-\omega ^2 x_f e^{2 r \tau }\nonumber\\
   &\qquad -r \omega  x_f e^{2
   r \tau } \sin (2 \tau  \omega )+\omega ^2 x_f - \omega  v_i e^{r \tau } \sin (\tau  \omega )+\omega  v_i e^{3 r \tau } \sin (\tau  \omega )\nonumber\\
   &\qquad+\omega ^2 x_i e^{r \tau
   } \left(e^{2 r \tau }-1\right) \cos (\tau  \omega )+r \omega  x_i e^{r \tau } \sin (\tau  \omega )+r \omega  x_i e^{3 r \tau } \sin (\tau  \omega )
\Bigl]
\end{align*}
\begin{align*}
a_3&=\Gamma e^{r \tau }\Bigl[r^2 x_f e^{r \tau } \sin (2 \tau  \omega )+\omega  v_f e^{r \tau }-\omega  v_f e^{3 r \tau }+r v_f e^{r \tau } \sin (2 \tau  \omega )\nonumber\\
   &\qquad+r \omega  x_f
   e^{3 r \tau }-r \omega  x_f e^{r \tau } \cos (2 \tau  \omega )-2 r^2 x_i e^{2 r \tau } \sin (\tau  \omega )-2 r v_i e^{2 r \tau } \sin (\tau  \omega )\nonumber\\
   &\qquad-\omega \left(e^{2 r \tau }-1\right) \cos (\tau  \omega ) \left(r x_i-v_i\right)-\omega ^2 x_i e^{2 r \tau } \sin (\tau  \omega )+\omega ^2 x_i \sin (\tau  \omega
   )
\Bigl]
\end{align*}
\begin{align*}
a_4&=\Gamma e^{r \tau }\Bigl[r^2 x_f \left(-e^{r \tau }\right)-r v_f e^{r \tau }+r e^{r \tau } \cos (2 \tau  \omega ) \left(r x_f+v_f\right)-\omega ^2 x_f e^{r \tau }\nonumber\\
   &\qquad+\omega ^2
   x_f e^{3 r \tau }+r \omega  x_f e^{r \tau } \sin (2 \tau  \omega )-\omega  v_i e^{2 r \tau } \sin (\tau  \omega )-\omega ^2 x_i \left(e^{2 r \tau }-1\right) \cos
   (\tau  \omega )\nonumber\\
   &\qquad-r \omega  x_i e^{2 r \tau } \sin (\tau  \omega )-r \omega  x_i \sin (\tau  \omega )+\omega  v_i \sin (\tau  \omega )
\Bigl]
\end{align*}

In the above equations, $\Gamma$ is
\begin{align}
\Gamma=\frac{1}{\omega ^2+\omega ^2 e^{4 r \tau }+2 r^2 e^{2 r \tau } \cos (2 \tau  \omega )-2 e^{2 r \tau } \left(r^2+\omega ^2\right)}
\label{eq:A2}
\end{align}

Choosing boundary condition as Eq.\ref{eq:BC2}, the classical action yields
\begin{equation}
S_c=S_c(x_f,v_f,x_i,v_i)=-\frac{1}{2}\sum_{I,J=1}^4 x_I M_{IJ} x_J
\nonumber
\end{equation}

From Eq. \ref{bcclassadd}, the classical action has the symmetry
\begin{eqnarray}
\mathcal{S}_c[x_f,v_f;x_i,v_i]=\mathcal{S}[x_i,-v_i;x_f,-v_f]
\end{eqnarray}
Hence under this boundary condition, the solution of matrix $M$ satisfies the following symmetry
\begin{align*}
&M_{11}=M_{44},\quad ~~M_{22}=M_{33}\nonumber\\
&M_{12}=-M_{34},\quad M_{13}=-M_{24}
\end{align*}
and the action can be simplified to
\begin{eqnarray}
\label{eq:action}
&&S_c(x_f,v_f,v_i,x_i)=-\frac{1}{2}M_{11}(x_i^2+x_f^2)-\frac{1}{2}M_{22}(v_i^2+v_f^2) \nonumber\\
&&+M_{12}(x_iv_i-x_fv_f)+M_{13}(x_iv_f-x_fv_i) -M_{14}x_ix_f-M_{23}v_iv_f
\end{eqnarray}
Solution of $M_{IJ}$ is listed below.
\begin{align*}
M_{11}&=\Gamma \Bigr[2 a  r \omega  \left(r^2+\omega ^2\right) \left(\omega  \left(e^{4 r \tau }-1\right)+2 r e^{2 r \tau } \sin (2 \tau  \omega )\right)\Bigl]\\
M_{12}&=\Gamma \Bigr[\omega ^2 e^{4 r \tau } \left(2 a  r^2-b \right)-2 r^2 e^{2 r \tau } \left(2 a  \omega ^2+b \right) \cos (2 \tau  \omega )\nonumber\\
&\qquad-\omega ^2 \left(b -2
   a  r^2\right)+2 b  e^{2 r \tau } \left(r^2+\omega ^2\right)
\Bigl]\\
M_{13}&=\Gamma \Bigr[4 a  r \omega  e^{r \tau } \left(e^{2 r \tau }-1\right) \left(r^2+\omega ^2\right) \sin (\tau  \omega )
\Bigl]\\
M_{14}&=\Gamma\Bigr[-4 a  r \omega  e^{r \tau } \left(r^2+\omega ^2\right) \left(r \left(e^{2 r \tau }+1\right) \sin (\tau  \omega )+\omega  \left(e^{2 r \tau }-1\right) \cos (\tau
    \omega )\right)
\Bigl]\\
M_{22}&=\Gamma \Bigr[-2 a  r \omega  \left(\omega  \left(-e^{4 r \tau }\right)+2 r e^{2 r \tau } \sin (2 \tau  \omega )+\omega \right)
\Bigl]\\
M_{23}&=\Gamma \Bigr[4 a  r \omega  e^{r \tau } \left(r \left(e^{2 r \tau }+1\right) \sin (\tau  \omega )-\omega  \left(e^{2 r \tau }-1\right) \cos (\tau  \omega )\right)
\Bigl]
\end{align*}

\subsection{Infinite time limit of classical action}
The infinite $\tau$ limit of matrix elements $M$ under the boundary condition $x_f, v_f, x_i, v_i$ are given by the following.
\begin{eqnarray}
\label{parainfinitie}
\lim_{\tau\rightarrow+\infty}M_{11}=2ra(r^2+\omega^2)~;~\lim_{\tau\rightarrow+\infty}M_{22}=2ra ~;~
\lim_{\tau\rightarrow+\infty}M_{12}=2r^2a-b~~~\\
\displaystyle\lim_{\tau\rightarrow+\infty}M_{13}=\displaystyle\lim_{\tau\rightarrow+\infty}M_{14}= \displaystyle\lim_{\tau\rightarrow+\infty}M_{23}=0\nonumber~~~~~~~~~~~~~~~~~~~
\end{eqnarray}

From the definition of the probability amplitude given in Eq. \ref{kerstasp}
\begin{equation*}
\mathcal{K}_S(x_f,v_f;x_i,v_i)=\bra{x_f,v_f}e^{-\tau H}\ket{x_i,v_i}=\bra{x_f,v_f}e^{Q/2}e^{-\tau H_0}e^{-Q/2}\ket{x_i,v_i}
\end{equation*}
yields, for infinite $\tau$ limit as given in Eq. \ref{kerlargetau}, the following
\begin{eqnarray}
&&\displaystyle\lim_{\tau\rightarrow+\infty}\mathcal{K}_S(x_f,v_f;x_i,v_i)\simeq e^{-\tau E_{00}} \bra{x_f,v_f}\Psi_{00}\rangle\langle \Psi_{00}^D\ket{x_i,v_i} +O(e^{-\tau(E_1-E_0)})\nonumber\\
&&~~~~= e^{-\tau E_{00}} \Psi_{00}(x_f,v_f)\Psi_{00}^D(x_i,v_i)=e^{-\tau E_{00}} \Psi_{00}(x_f,v_f)\Psi_{00}(x_i,-v_i) \nonumber
\end{eqnarray}
Hence, from Eq. \ref{kerstaspinfinitie}, the infinite limit of transition amplitude is the product of vacuum state $\Psi_{00}^D(x_i,v_i)$ and $\Psi_{00}(x_f,v_f)$ and the classical solution yields
\begin{align}
&\displaystyle\lim_{\tau\rightarrow+\infty}K(x_f,v_f,x_i,v_i)
\label{kerstaspinfinitie}
=\displaystyle\lim_{\tau\rightarrow+\infty} \mathcal N' e^{S_c(x_f,v_f,x_i,v_i)}\nonumber\\
&~~~~~~~~~~~~~~~~~~~~~~~~~~~~~~= e^{-\tau E_{00}} \Psi_{00}(x_f,v_f)\Psi_{00}(x_i,-v_i)
\end{align}

The infinite time classical action has precisely the factorization of the initial and final boundary values as given in Eq. \ref{kerstaspinfinitie}. Therefore, the vacuum state, obtained from Eqs. \ref{eq:action} and \ref{parainfinitie} and, defining $x_f=x,v_f=v$, yields the following
\begin{align}
\label{vacpathint}
\Psi_{00}(x,v)&=\displaystyle\lim_{\tau\rightarrow+\infty} \mathcal N \exp\left(-\frac{1}{2}M_{11}x^2-\frac{1}{2}M_{22}v^2-M_{12}xv\right)\nonumber\\
&=\mathcal N \exp\Bigl(-ra^2(r^2+\omega^2)x^2-r a v^2-(2r^2a-b)xv\Bigr)
\end{align}
where $\mathcal N$ is fixed by normalizing $\Omega(x,v)$.

To make connection with the earlier parametrization, rewrite the Lagrangian in $\omega_1$ and $\omega_2$ parametrization (the only difference with the parametrization given in Eq. \ref{lagpot} is in the overall factor of $a$ instead of $\gamma$)
\begin{align}
\mathcal L&=-\frac{1}{2}a \left(\ddot x^2+2 \frac{b}{a} \dot x^2+\frac{c}{a} x^2\right)\nonumber\\
&=-\frac{1}{2}a \Bigl(\ddot x^2+(\omega_1^2+\omega_2^2)\dot x^2 +(\omega_1^2\omega_2^2)x^2\Bigr)
\end{align}
where $\omega_1$ and $\omega_2$ are\footnote{Note that $\omega_1$ and $\omega_2$ in Eq. \ref{omega1omega2} have a different parametrization, given by
\begin{eqnarray*}
\omega_1=\frac{1}{2\sqrt{\gamma}}\left(\sqrt{\alpha+2\sqrt{\gamma\beta}}+\sqrt{\alpha-2\sqrt{\gamma\beta}}\right)~~;~~
\omega_2=\frac{1}{2\sqrt{\gamma}}\left(\sqrt{\alpha+2\sqrt{\gamma\beta}}-\sqrt{\alpha-2\sqrt{\gamma\beta}}\right)\nonumber
\end{eqnarray*}}
\begin{align*}
\omega_1=\frac{1}{\sqrt{2a}}\left(\sqrt{b+\sqrt{ac}} +\sqrt{b-\sqrt{ac}}\right)\\
\omega_2=\frac{1}{\sqrt{2a}}\left(\sqrt{b+\sqrt{ac}} -\sqrt{b-\sqrt{ac}}\right)
\end{align*}

The definition variables $r$ and $\omega$ in terms of $\omega_1$ and $\omega_2$ is given by
\begin{eqnarray*}
\omega_1=r+i \omega~~;~~\omega_2=r-i \omega
\end{eqnarray*}
The two parameterizations have the following relationship
\begin{eqnarray}
\label{defrw}
r^2+\omega^2&=\omega_1^2\omega_2^2~~;~~r=\frac{1}{2}(\omega_1+\omega_2)~~;~~b=\frac{a}{2}(\omega_1^2\omega_2^2)
\end{eqnarray}
Eqs. \ref{vacpathint} and \ref{defrw} hence yield -- replacing $a$ by $\gamma$ to conform to the notation given in Eq. \ref{omega1omega2} -- the vacuum state given in Eq. \ref{vacuumhigh}, namely
\begin{eqnarray}
\label{vacaction}
\Psi_{00}(x,v)=\mathcal N \exp\left(-\frac{\gamma}{2}(\omega_1+\omega_2)\omega_1\omega_2x^2-\frac{\gamma}{2}(\omega_1+\omega_2) v^2-\gamma\omega_1\omega_2 xv\right) 
\end{eqnarray}

The definition of the evolution kernel  $\mathcal{K}_S(x_f,v_f;x_i,v_i)=\bra{x_f,v_f}e^{-\tau H}\ket{x_i,v_i}$ given in Eq. \ref{kerstasp} is seen to be correct since the  evolution kernel obtained from the classical solution also gives the same result for the vacuum state as given by the Hamiltonian in Eq. \ref{vacuumhigh}. 

\bibliography{../../master_references_all}\label{refs}
\bibliographystyle{plain}
\end{document}